# Search Disaster Victims using Sound Source Localization

**Abhish Khanal**
Institute of Engineering, Pulchowk Campus
072bex402@ioe.edu.np

**Deepak Chand**
Institute of Engineering, Pulchowk Campus
072bex416@ioe.edu.np

**Prakash Chaudhary**
Institute of Engineering, Pulchowk Campus
072bex429.prakash@pcampus.edu.np

**Subash Timilsina**
Institute of Engineering, Pulchowk Campus
072bex442.subash@pcampus.edu.np

**Sanjeeb Prasad Panday**
Institute of Engineering, Pulchowk Campus
sanjeeb@ioe.edu.np

**Aman Shakya**
Institute of Engineering, Pulchowk Campus
aman.shakya@ioe.edu.np

**Rom Kant Pandey**
Sanothimi Campus, Tribhuvan University
romkant@gmail.com

**ABSTRACT**

Sound Source Localization (SSL) are used to estimate the position of sound sources. Various methods have been used for detecting sound and its localization. This paper presents a system for stationary sound source localization by cubical microphone array consisting of eight microphones placed on four vertical adjacent faces which is mounted on three wheel omni-directional drive for the inspection and monitoring of the disaster victims in disaster areas. The proposed method localizes sound source on a 3D space by grid search method using Generalized Cross Correlation Phase Transform (GCC-PHAT) which is robust when operating in real life scenario where there is lack of visibility. The computed azimuth and elevation angle of victimized human voice are fed to embedded omni-directional drive system which navigates the vehicle automatically towards the stationary sound source.

**Keywords**

Sound Source Localization (SSL), Omni-Directional Drive, Disaster Victim, Generalized Cross Correlation Phase Transform (GCC-PHAT).

**INTRODUCTION**

Sound source localization means finding the position of the sound source in the environment. Localization of sound source through embedded systems can be advantageous since it can be placed on various robots and devices to know the coordinates of an object . For example, you will be able to hear where the sound is coming from and will be able to move in the direction where the sound is coming to respond. In the modern time, where automation technologies are developing drastically, sense of hearing can provide great benefits to robots and other devices while rescuing the disaster victims during the time of crisis (Sun et al. 2011; Park et al. 2017). The work presented in this paper aims toward design of stationary sound source localization, especially human voices to locate them during the time of disaster where there is lack of visibility.The system is designed and deployed on an omni-directional ground vehicle using cubical microphone array which navigates to the place where the sound comes from.
Various measures has been researched in audition system for robot which are inspired by human audition system and rely on Time difference on arrival (TDOA). Gala et al. (2018) designed three-dimensional sound source localization for unmanned ground vehicles with a self-rotational two-microphone array. Cho et al. (2009) designed sound source localization for robot auditory systems. DiBiase et al. (2001) designed Robust localization in reverberant rooms. Similarly, Beh et al. (2010) designed sound source separation by using matched beamforming and time-frequency masking and Padois (2018) designed acoustic source localization based on the generalized cross-correlation and the generalized mean with few microphones. Application of the methods to control vehicles in real environment where in-band noises are present have shown difficulties in transferring theoretical methods to a real life scenario.

Although development of human like auditory system and implementing on vehicle to have exact senses as human auditory system is challenging, increasing the number of microphones in the array shows significant improvement in robust localization (Valin et al. 2003). Adding techniques to filter all the environment noise can increase the robustness and accuracy of the system to transfer these methods to real life scenario where more than one sources are present.

In this regard, this paper presents sound source localization using grid search method. The detail of this method is described in the section sound source localization using grid search method. The hardware implementation of cubical microphone array and embedding to robotic platform is explained in the section hardware setup. This is followed by the implementation result in experimental result section.

**RELATED WORKS**

In the field of sound source localization, there are various techniques used for robust localization. Gala et al. (2018) has done three-dimensional sound source localization for ground unmanned vehicle with rotational two microphone array. Here, two microphones are used to calculate the relative time difference of arrival of the signal to the microphones and hence the direction of the signal is calculated.

Cho et al. (2009) has located sound source using steered response power phase transform (SRP-PHAT). Here the generalized cross correlation phase transform (GCC-PHAT) of all the possible combination of microphone is added together to give relative position of sources in all the direction.

DiBiase et al. (2001) has done work on robust localization in reverberant rooms by using concept of time difference of arrival. Beh et al. (2010) have done work on sound separation using matched beamforming and time-frequency masking. Beamforming is a technique which is used to calculate the direction of arrival (DOA) of sound source by adding shifted version of sound source to a reference sound source. By adding all the shifted version of sound source, beam of high amplitude gives the direction of the sound source.

Padois (2018) has done work on Acoustic source localization based on the generalized cross correlation and generalized mean. Valin et al. (2003) has done work in Robust sound source localization using microphone array on mobile robot. This paper shows increasing the number of microphones performs robust localization with a trade-off of computational cost. This paper shows implementation of Cubical microphone array configuration to localize sound source using GCC-PHAT method and mounting it on a robotic platform to move towards the direction of sound source.

Grondin and Michaud (2019) has done work on Lightweight and optimized sound source localization and tracking methods for open and closed loop microphone array configurations. The main feature used in this approach is SRP-PHAT which gives probability of sound source on 3D space and Kalman filtering for localizing exact position of sound source from predicted sound source.

Doroftei and Stirbu (2010) and Tajti et al. (2014) has done work on Designing Omni-directional mobile robot. This paper presents all the mathematics for inverse kinematics used in this report to drive omni-directional vehicle.

Sun et al. (2011) primarily focuses on different from the light, ultrasonic sound and infrared signals, sound signals which can bypass obstacle and the development of the auditory navigation system for fire and earthquake rescue robot in unknown environment using time difference of arrival and speech recognition technology. Park et al. (2017) address fundamental problems with ground robots for disaster response and recovery with the design of robot and human control.

This paper proposes searching disaster victims using GCC-PHAT and grid search algorithm with combination of unique microphone configuration for describing the sound source around its periphery with less error and hence providing almost accurate position of the source.

**SOUND SOURCE LOCALIZATION USING GRID SEARCH METHOD**

**Time Difference of Arrival (TDOA)**

This is basic method for sound source localization (DiBiase et al. 2001). In this method if microphones $m1$ and $m2$ are $d$ distance apart, given that $c$ is the velocity of sound, then time difference of arrival (TDOA) from one microphone to other is given by (1):

$$\tau = \frac{d\sin\theta}{c} \tag{1}$$

Hence direction of arrival $\theta$ can be measured if TDOA $\tau$ is known. TDOA $\tau$ can be measured from cross correlation of two signals from (2) and (3):

$$R_d = \sum_{n=0}^{N-1} x_1[n] x_2[n+k] \tag{2}$$

$$\tau = \frac{1}{f_c} argmax(R_d[k]) \tag{3}$$

for more robust cross correlation, we have used Generalized Cross Correlation Phase Transform (GCC-PHAT).

## Generalized Cross Correlation Phase Transform (GCC-PHAT)

The correlation or cross power spectrum of two signals can also be found through frequency domain as:

$$R_{12}(\tau) = X_1(\omega) \overline{X_2(\omega)} \tag{4}$$

taking inverse Fourier transform, we get cross power spectrum as:

$$r_{12}(\tau) = \frac{1}{2\pi} \int_{-\infty}^{\infty} X_1(\omega) \overline{X_2(\omega)} e^{j\omega\tau} d\tau \tag{5}$$

Since the sound signal may contain its reverberated part, only phase information is needed when carrying out convolution. So magnitude $\left|X_1(\omega)\overline{X_2(\omega)}\right|$ is divided from $R_{12}(\tau)$ for better performance and then inverse Fourier transform is performed. Thus we get $r_{12}(\tau)$ as:

$$r_{12}(\tau) = \frac{1}{2\pi} \int_{-\infty}^{\infty} \frac{X_1(\omega) \overline{X_2(\omega)}}{\left|X_1(\omega) \overline{X_2(\omega)}\right|} e^{j\omega\tau} d\tau \tag{6}$$

## Grid Search Method Using GCC-PHAT

For grid search method, 3D spherical grid is constructed with a constant radius as shown in the Fig. 1 with visualizing microphone array in the center (Grondin and Michaud 2019).

After the construction of grid in 3D space with required resolution, the exact estimation of the source is done. For this estimation,

- We assigned the number of source, interpolation resolution and microphone coordinates. The distance between each microphone pair combination is calculated then the calculation of difference in distance from a point in grid to microphone pair is followed by the calculation of time shift for that point.

- We computed GCC-PHAT for each grid point combination from each microphone pair and added all of them. We used short time fast fourier transform (ST-FFT) in GCC-PHAT for less computational complexity.

- Finally we searched peaks from numerous peak. Group of peaks shows multiple sound source.

For reducing computational complexity, peak search is based on array indices of peaks. Indices of peaks gives exact azimuth and elevation in 3D grid given that each point in 3D grid is arranged in one dimensional array as shown in Fig. 2.

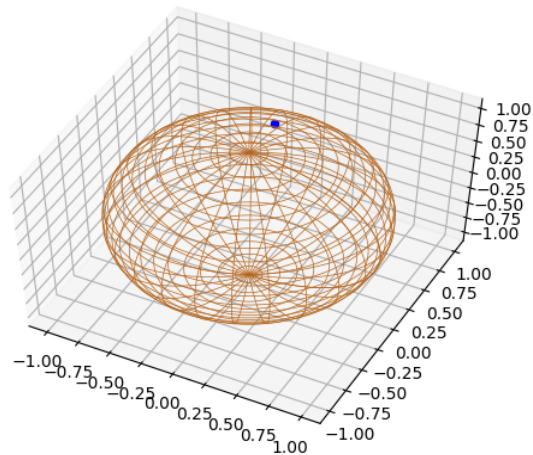

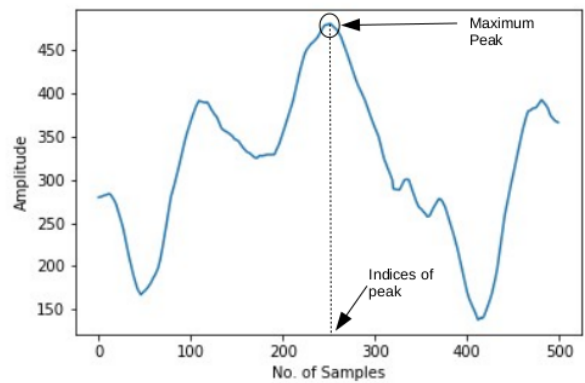

**Figure 1. Grid construction in 3D space**

**Figure 2. Maximum peak and indices of sum of GCC-PHAT**

## HARDWARE SETUP

### Cubical Microphone Array Configuration

The cube with dimension 15cm∗15cm∗15cm was designed in Solidworks and aluminum frame was used to fabricate the design because of its lightweight. The design made in Solidworks is shown in Fig. 3.

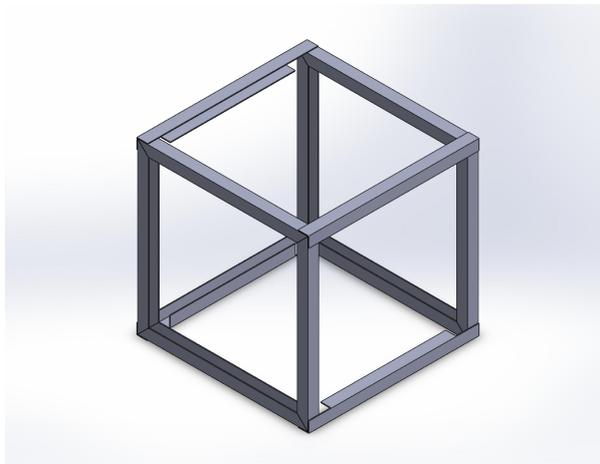

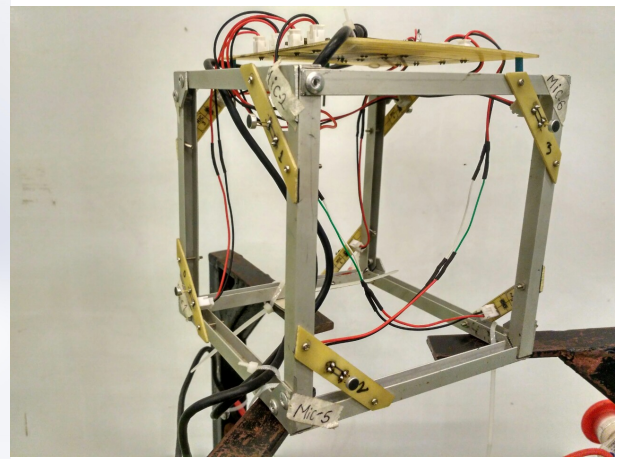

**Figure 3. Design of cube**

**Figure 4. Cubical arrangement of microphones**

Microphones are arranged in a cube with the inter microphone distance 15cm across the diagonals and placed in vertical adjacent faces as shown in Fig. 4.

The Microphones were dismantled from SONY PS2 eye and reconnected to the circuits by designing a new circuit on KiCAD for SONY PS2 eye circuit and Sony ps2 eye microphones as shown in Fig. 5 and Fig. 6.

The coordinates of microphones is taken from the center of the cube and this coordinate is required to calculate the distance of each microphone to the individual grid points in the 3D space. The omni-directional microphones are from SONY PS2 eye device. The data is received via USB protocol which is default for SONY PS2 eye.

The microphones can be placed in any positions along the cube with equal inter microphone distance as shown in Fig. 7 and grid search method can be used to estimate the position of the sound source.

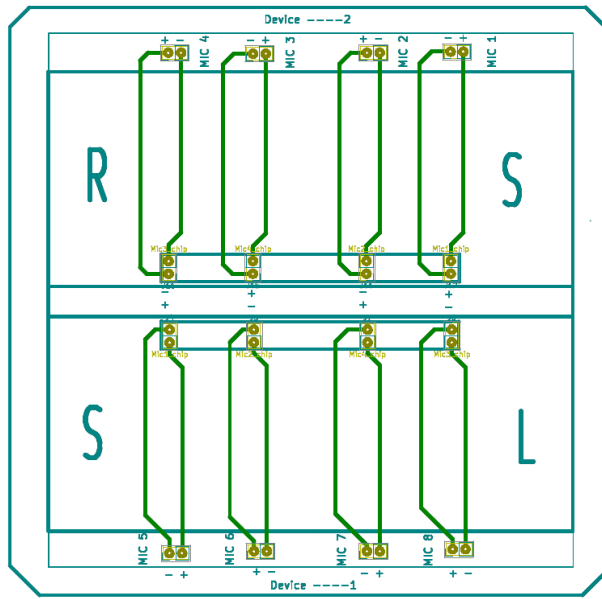

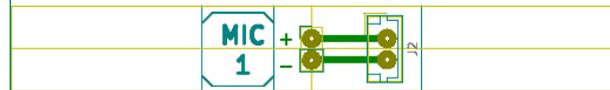

Figure 6. Design of microphone circuit

Figure 5. Design of circuit board for SONY PS2 eye

**Three wheel Omni-directional drive**

Cubical microphone array is mounted on Omni-directional drive with three omni-directional wheels arranged in $120^o$. The inverse kinematics of three wheeled omni-directional drive is implemented as reviewed in Doroftei and Stirbu (2010) and Tajti et al. (2014).The base of the omni-directional drive is shown in Fig. 8.

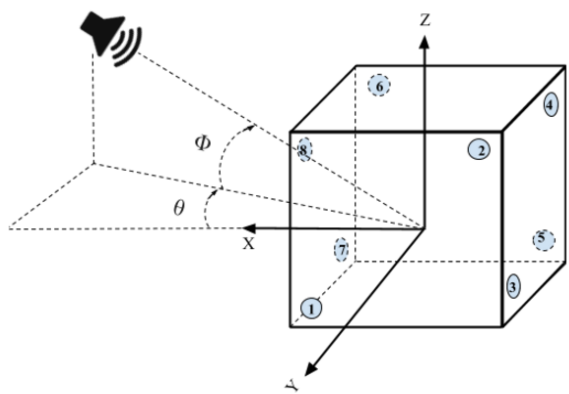

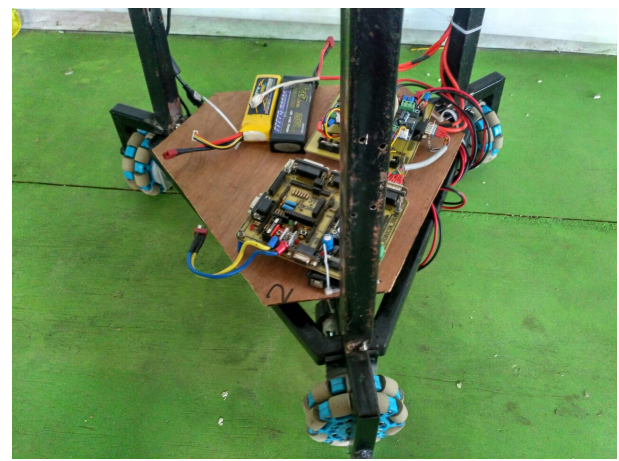

Figure 7. Microphone geometry in Cube

Figure 8. Three wheeled omni-directional drive

The motors of omni-wheels are controlled with Atmega2560 micro-controller which receives direction from raspberry pi.

**Proposed system**

The data from cubical microphone array is received in Raspberry Pi 3 via USB protocol and using grid search method, Azimuth and elevation of the sound source is calculated. Through Serial port, Azimuth angle is sent to Embedded Atmega2560 micro-controller which drives the omni-directional drive to the location where the sound comes from.

The complete structure of cubical microphone array embedding with omni-directional drive using raspberry pi is shown in Fig 10.

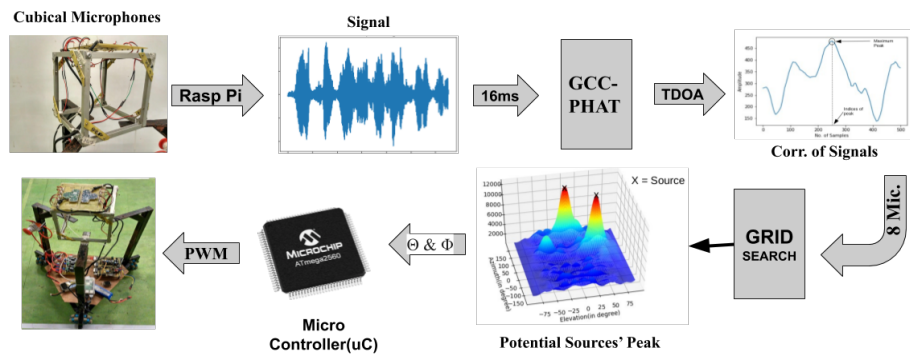

Figure 9. flowchart of proposed system

## EXPERIMENTAL RESULT

An experimental setup was conducted in which omni-directional vehicle was placed and sound source of 62.5 dB was moved with a constant azimuth and elevation of $-4^o$ and $-45^o$ respectively. 50 cm from the vehicle, azimuth and elevation data was taken for each increase in 10 cm upto 300 cm and the error in azimuth was measured and plotted with respect to distance. The result is shown in Fig. 11.

As the distance increases, Mean squared error increases gradually which is due to less sensitivity of microphones. The mean squared error can be decreased by using more sensitive microphones.

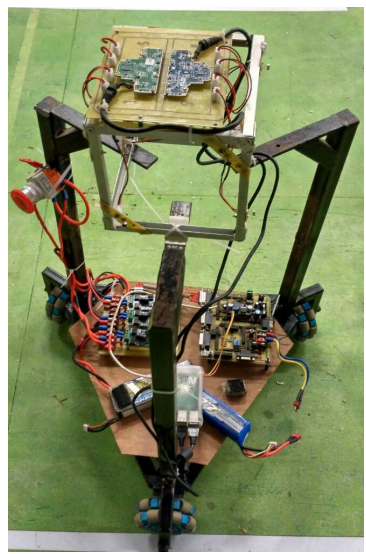

Figure 10. Structure of three wheeled omni-directional vehicle with cubical microphone array

A setup was conducted where omni-directional vehicle was placed in center and sound source of 62.5 dB was varied in azimuth and elevation. The average azimuth and elevation recorded separately are shown in Fig.13,14,15,16 as box plot and filled in Table II.

The azimuth and elevation angle were measured by keeping elevation constant when measuring azimuth and vice-versa. The source was kept at 150 cm distance from the omni-directional vehicle. Average error of $1^o$ on azimuth and on elevation was observed over 150 cm range. Three sources were placed at 200 cm radius from omni-directional vehicle. Source 1 at azimuth= $-135^o$ and elevation= $35^o$, Source 2 at azimuth= $90^o$ and elevation= $75^o$ and Source 3 at azimuth= $145^o$ and elevation= $60^o$ were placed. The position measured is shown in Fig. 17 with peaks in surface plot.

There are several peaks in the surface plot which represents source as shown in Fig. 17 but the maximum three peaks denote three sources. For more sources, more peaks has to be observed. The position of source 1 was found to be azimuth= $-130^o$, elevation= $32^o$. The position of source 2 was found to be azimuth= $87^o$, elevation= $73^o$. The position of source 3 was found to be azimuth= $144^o$, elevation= $60^o$. The azimuth angle was fed to omni-directional

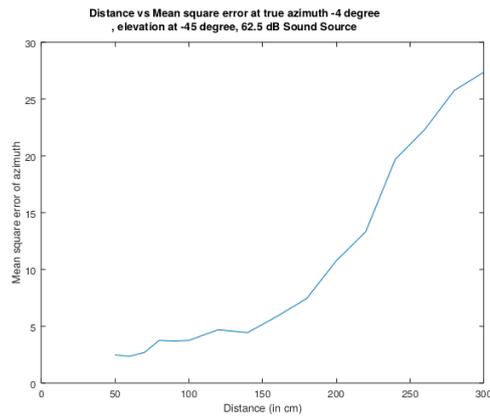

Figure 11. Distance vs Mean Squared Error (MSE)

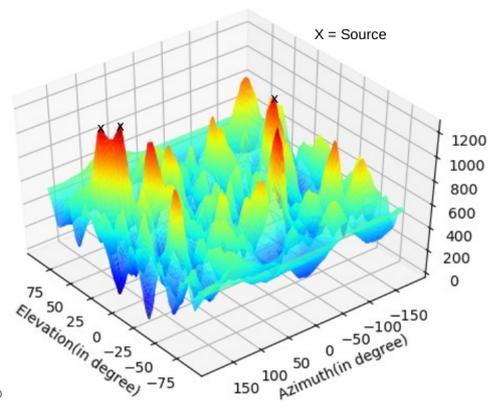

Figure 12. Surface plot of peaks obtained from various sources

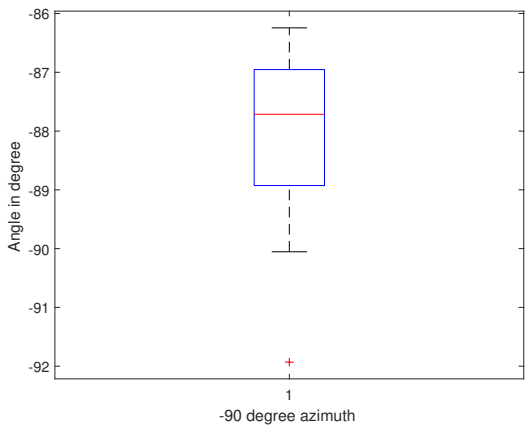

Figure 13. Box plot for azimuth at $-90^o$

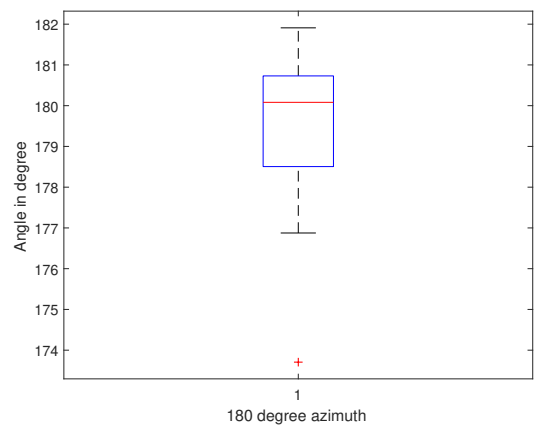

Figure 14. Box plot for azimuth at $180^o$

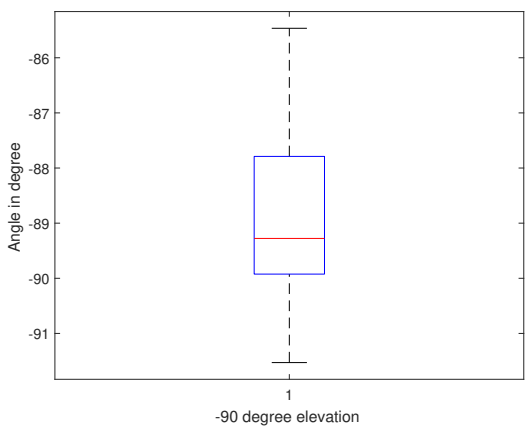

Figure 15. Box plot for elevation at $-90^o$

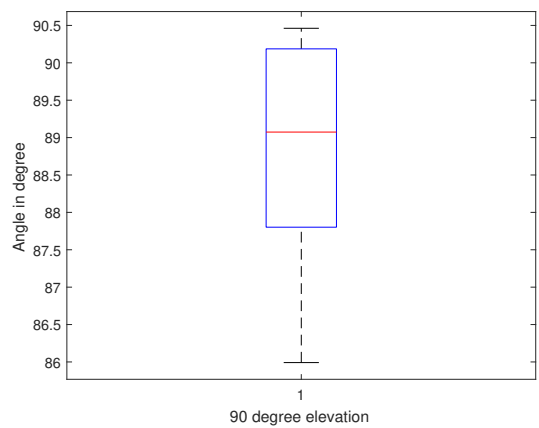

Figure 16. Box plot for elevation at $90^o$

Table 1. Measured Elevation and Azimuth

| True Azimuth (degree) | True Elevation (degree) | Measured Azimuth (degree) | Measured Elevation (degree) |
|---|---|---|---|
| 180 | -90 | 179 | -88 |
| 45 | -45 | 44 | -47 |
| 0 | 0 | 2 | -2 |
| -45 | 45 | -44 | 44 |
| -90 | 90 | -88 | 89 |

vehicle to move towards the direction of sound source. The vehicle moves towards the most probable sound source although the system is able to detect multiple sources. The vehicle tracks the source with highest energy even if two or more sources are present.

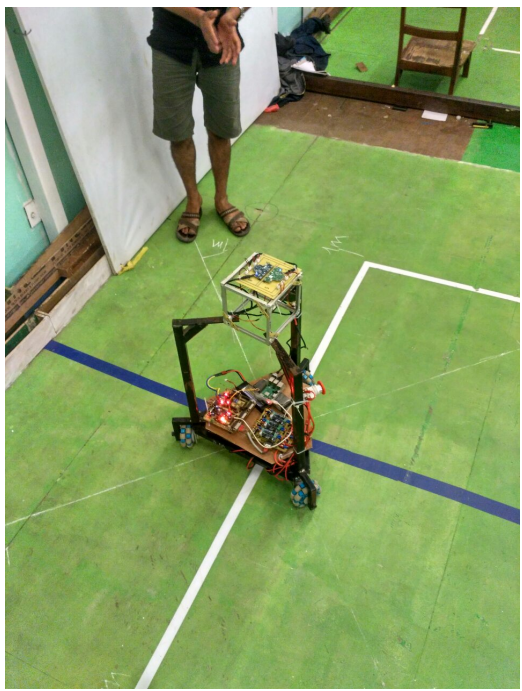
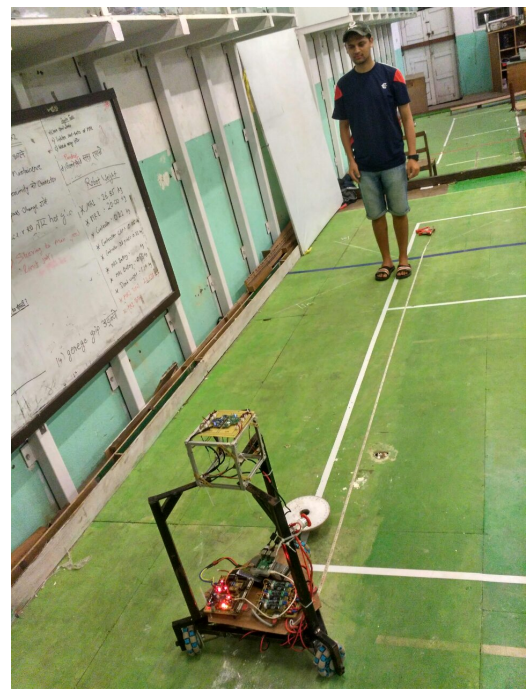

(a) Experimental setup for tracking sound source    (b) Experimental setup for measuring distance vs accuracy

Figure 17. Photos during experiment

## CONCLUSION AND FUTURE WORK

In this paper, we present an approach of localizing sound source using cubical microphone array and implementing it on an omni-directional drive for inspection and monitoring the disaster area which helps to rescue the human casualties. The localization system predicts both the azimuth and elevation angle of human voice with less error. Certain error was observed in the system which was due to microphone sensitivity and also trade-off between resolution of grid search method and computational cost. So, there are still lots of room for improvement. We would implement the following future works to make our system more robust.

- The land vehicle only follows the sound source in 2D plane instead our system predicts both azimuth and elevation, so we plan to deploy our system on Unmanned Aerial Vehicle (UAV).

- Different noise such as environmental noise degrade the performance of the system, for this different advance filtration techniques can be used.

- By using the sensitive microphones we can predict the source quite accurately.

- Single sound source is tracked from multiple sources discovered, sound source are prioritize according to the energy of the source. Further development can be done to prioritize the victims.

**ACKNOWLEDGEMENT**

This work has been supported by the University Grants Commission, Nepal under a Collaborative Research Grant (UGC Award No. CRG-74/75-Engg-01) for the research project "Establishment of a Disaster Telecommunications Research and Educational Facility Advancing a Scientifically Sound Disaster Telecommunication Infrastructure and Processes in Nepal".